\documentclass[reprint,superscriptaddress,amsmath,amssymb,aps,prl]{revtex4-1}
\usepackage{hyperref}
\usepackage{graphicx}
\usepackage{mathrsfs}  
\usepackage{xcolor}
\usepackage{dcolumn}
\usepackage{physics}
\usepackage{bm}
\usepackage{amsmath}
\usepackage[all]{xy}
\usepackage{ulem} 
\usepackage[utf8]{inputenc}

\newcommand{\floor}[1]{\lfloor #1 \rfloor}
\makeatletter
\newcommand*{\Relbarfill@}{\arrowfill@\Relbar\Relbar\Relbar}
\newcommand*{\xeq}[2][]{\ext@arrow 0055\Relbarfill@{#1}{#2}}
\makeatother

\newcommand{\hH}[0]{\hat{H}}
\newcommand{\hPsid}[0]{\hat{\Psi}^\dagger}
\newcommand{\hPsi}[0]{\hat{\Psi}}

\newcommand{\ssec}[1]{\emph{#1}.---}
\newcommand{\sign}{\mathop{\rm sign}\nolimits}

\newcommand{\meanv}[1]{\langle #1\rangle}

\begin{document}
\title{
Fermionizing the ideal Bose gas via topological pumping
%Topological pumping as a method of creating a strongly correlated ideal Bose gas
%\kp{Topological pumping as a method of creating a fermionized state of a 1D ideal Bose gas}
}

\author{Maciej Marciniak}
\affiliation{
Center for Theoretical Physics, Polish Academy of Sciences, Al. Lotnik\'{o}w 32/46, 02-668 Warsaw, Poland
}

\author{Grigori E. Astrakharchik}
\affiliation{Departament de Física, Universitat Politècnica de Catalunya, E-08034 Barcelona, Spain}

\author{Krzysztof Pawłowski}
\affiliation{
Center for Theoretical Physics, Polish Academy of Sciences, Al. Lotnik\'{o}w 32/46, 02-668 Warsaw, Poland
}

\author{Bruno Juliá-Díaz}
\affiliation{Departament de Física Quàntica i Astrofísica, Facultat de Física, Universitat de Barcelona, 08028 Barcelona, Spain}
\affiliation{Institut de Ciències del Cosmos, Universitat de Barcelona, Martí i Franquès 1, 08028 Barcelona, Spain}

\date{\today}

\begin{abstract}
%We theoretically predict that topological pumping -- a recently realized experimental technique -- can be used to fermionize an ideal Bose gas in one dimension. Specifically, the momentum distribution of the non-interacting homogeneous bosons after $n$ pumping cycles becomes equal to that of ideal fermions but at $2n$-times higher density. The coherence function of the pumped states decays exactly as in the Fermi gas whereas density-density correlations still exhibit bosonic enhancement. Our study opens pathways to new methods for generating strongly correlated quantum states.
We investigate the coherence and correlations of many-body states appearing in topological pumping in a one-dimensional Bose gas. By analyzing the system at zero and infinite interaction strengths, we reveal a rescaling of momentum distributions accompanied by a self-similar behavior in first- and second-order correlation functions. In excited states of non-interacting bosons, the momentum distribution shows a comb-like structure similar to that of non-interacting fermions but at a higher density. This is mirrored by Friedel oscillations in the one-body density matrix. At the same time, the density-density correlations still exhibit the bosonic enhancement. Our work illustrates how topological pumping induces a nontrivial mapping between bosonic and fermionic correlations.
\end{abstract}
\maketitle

Although bosons and fermions differ fundamentally, there exists a physical system in which they can be related via an elegant mapping. Such a connection occurs in the one-dimensional (1D) quantum many-body system with short-range interactions modeled via the Lieb-Liniger (LL) model~\cite{Lieb1963May, Lieb1963Excited}. In a regime of infinitely strong repulsion between bosons, in the so-called Tonks-Girardeau regime (TG), known also as the fermionization limit~\cite{Girardeau2012Dec}, all bosonic eigenstates can be mapped to the eigenstates of non-interacting fermions. Eigenenergies, and all density-based correlations, including the second-order correlation functions, are identical for both systems. Physically, the similarities have the origin in the strong short-range repulsion preventing bosons from occupying the same position, mimicking the effect of the Pauli principle for fermions.

Another peculiarity of the 1D system emerges when the sign of the interaction strength is rapidly reversed, from strong repulsion to strong attraction between atoms, for instance by crossing the confinement-induced resonances~\cite{Olshanii1998Aug,Haller2010Apr} or Feshbach resonances~\cite{Chin2010Apr}. The bosonic state transition to the so-called super-Tonks-Girardeau (sTG) state which remains metastable despite strong attraction between atoms, as shown theoretically~\cite{Astrakharchik2005Nov,Batchelor2005Oct,Chen2010Mar,Girardeau2012Dec} and 
experimentally~\cite{Haller2009Sep, Kao2021Jan}.
%%%%%%%%%%% FIGURE 1: GRAPHICAL ABSTRACT  
\begin{figure}[ht!]
\centering
\includegraphics[width=\columnwidth]{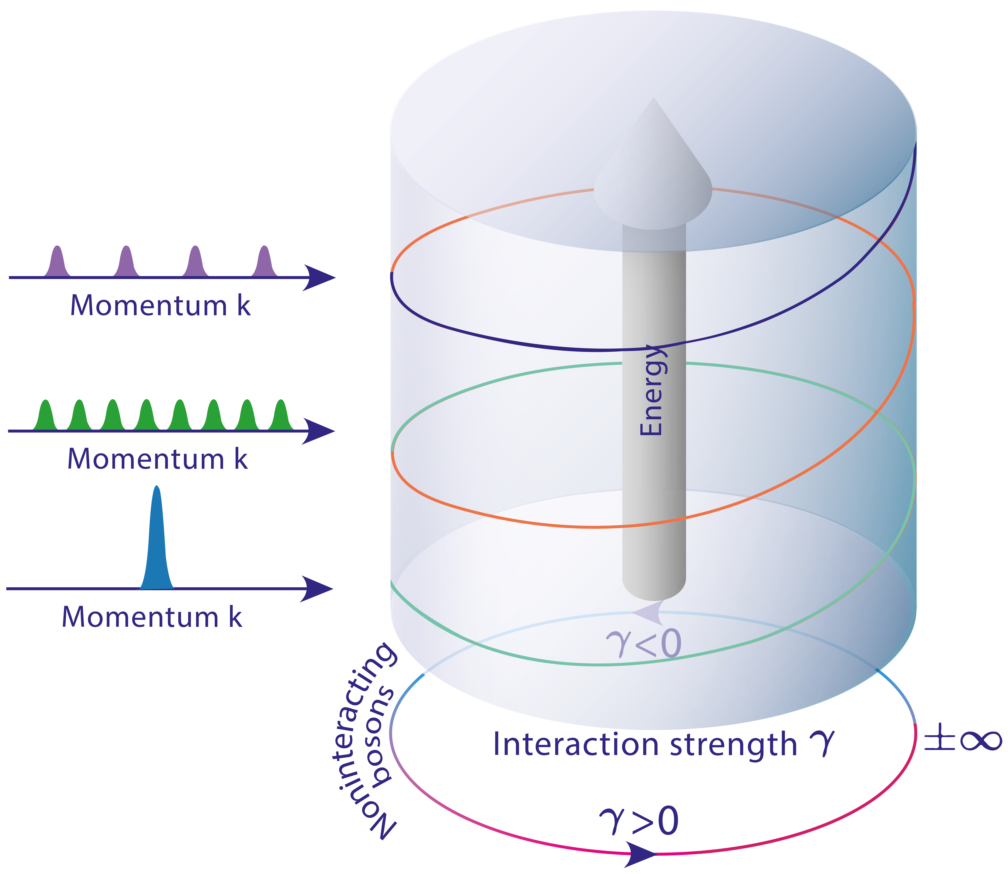}
\caption{Illustration of the ``tower of states'' emerging during cyclical changes in the interaction parameter $\gamma$, shown as a circle on the bottom in a stereographic representation with $\pm\infty$ being a single point. Left panels: the characteristic comb-like momentum distribution in the appearing excited states of an ideal Bose gas (compare with Fig.~\ref{fig:man_LL_momentum}). The colors of the helical line correspond to different cycles (see Fig.~\ref{fig:man_quasimomenta}b).\label{fig:man_graphical_abstract}}
\end{figure}
%%%%%%%%%%% END OF FIGURE 1
As demonstrated in Ref.~\cite{Kao2021Jan}, by repeatedly crossing the resonance in the 1D interactions, the system can be adiabatically excited, following the pumping protocol, as schematically shown in Fig.~\ref{fig:man_graphical_abstract}. In the pumping protocol, the interaction strength is first smoothly increased from $\gamma=0$ (non-interacting bosons) up to $\gamma\to+\infty$ (fermionization limit), maintaining the gas in the ground state. The subsequent quench of the interaction strength pushes the gas to the metastable sTG regime with $\gamma\to-\infty$. This quench is followed by the next stage of adiabatic dynamics, a smooth increase of interactions back to $0$ to end the first cycle. Despite interaction quenches, the changes of states are always continuous. The subsequent cycles pump the system to higher and higher energy eigenstates.

The sTG state, a key element of the pumping protocol, has been experimentally realized in ultracold trapped atoms already in 2009~\cite{Haller2009Sep}. Still, pumping was not possible then due to instabilities at intermediate attractive interaction strength~\cite{Chen2010Mar}. These difficulties were overcome recently by using Dysprosium atoms with dipole-dipole repulsion~\cite{Kao2021Jan,Li2023Jun,Yang2023Aug}, preventing instabilities and enabling the experimental realization of a few cycles.

In this Letter, we demonstrate that pumping can be used to achieve states of {\it non-interacting bosons} that resemble ground-state of fermions. In particular, the states emerging at $n$-th pumping cycle have single bosons occupying every $2n$-th energy level, up to Fermi energy multiplied by $2n$. In the special case of the box with periodic boundary conditions, the momentum distribution of these bosonic states is exactly the same as that of non-interacting fermions, but confined in a $2n$ times shorter box,  see Fig.~\ref{fig:man_graphical_abstract}. Consequently, the first-order correlation function of these non-interacting states is at short-length scales, similar to the correlation function of fermions.
%Their kinetic energy is exactly $(2n)^2$ times larger than in the ground state of non-interacting fermions. 

We start by discussing the Lieb-Liniger model to introduce the notation and key concepts. Then, we characterize the states emerging during topological pumping in terms of Lieb's quasi-momenta, followed by an analysis of inter-particle correlations and the corresponding equation of state.

\begin{figure}
\centering
\includegraphics[trim= 1.5cm 0cm 0cm 0cm, width=\columnwidth]{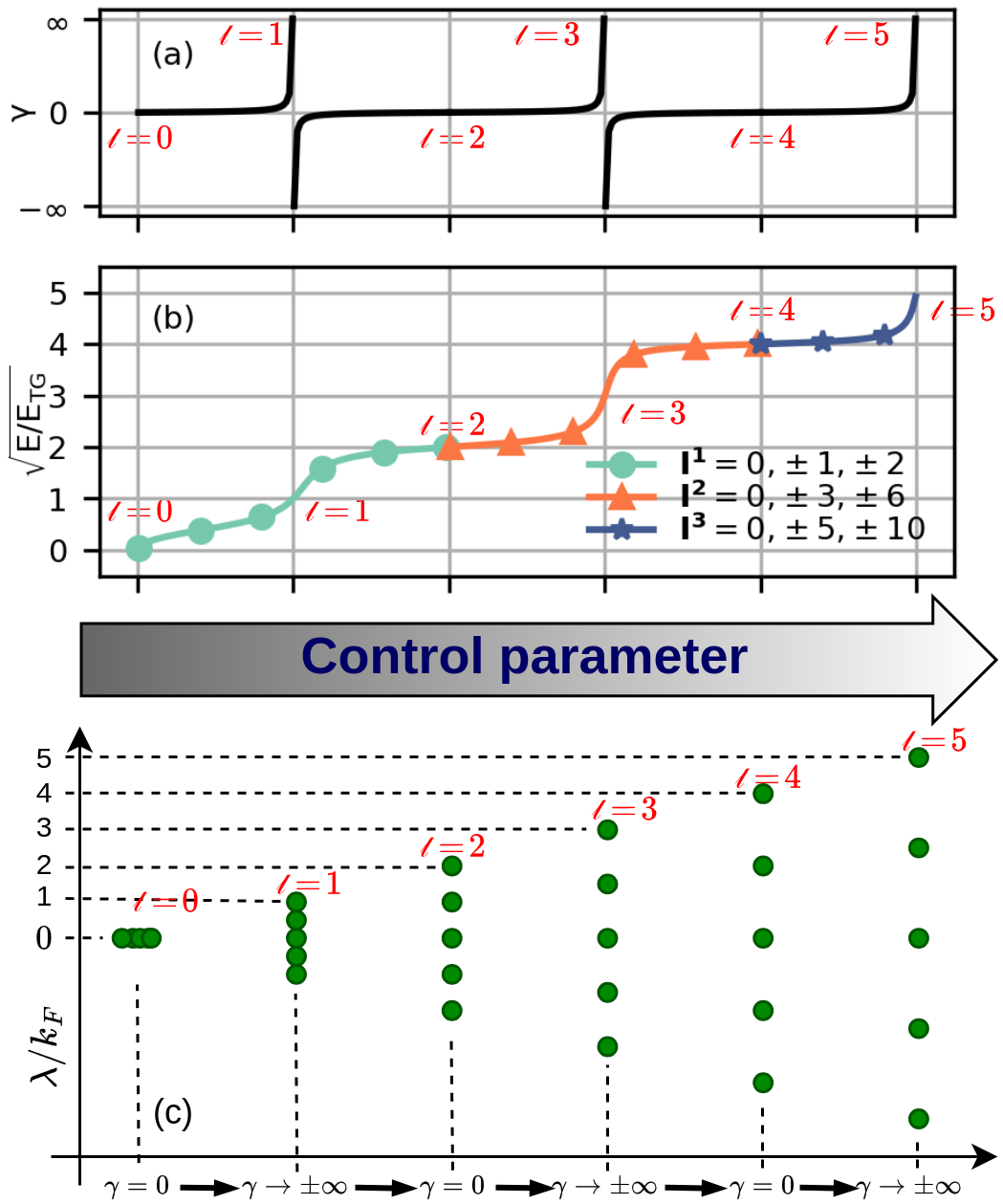}
\caption{
Illustration of the topological pumping process, resembling experiment~\cite{Kao2021Jan}. We present the evolution throughout the pumping of
(a) Lieb-Liniger parameter $\gamma$
(b) square root of the energy $\sqrt{E/E_{\rm TG}}$ of states appearing during the pumping and (c) their  Lieb's quasi-momenta $\lambda$, taking $N=5$ atoms for the example. The corresponding Bethe numbers $\bm{I}$ are shown in the panel (b). Here $E_{\rm TG}$ is the energy of the Tonks-Girardeau ground state, and $k_{\rm F} = \pi \rho$. 
\label{fig:man_quasimomenta}}
\end{figure}

%%%%%%%%%%%%%%%%%%%%%%%%%%%%%%%%%%%%%%%%%%%%%%%
\ssec{Box with periodic boundary conditions}%%%
%%%%%%%%%%%%%%%%%%%%%%%%%%%%%%%%%%%%%%%%%%%%%%%
First, we consider a gas of $N$ indistinguishable bosons having linear density $\rho = N/L$ and confined in a one-dimensional box of size $L$ with periodic boundary conditions (PBC). 
We assume that atoms interact via a contact pseudopotential.
Such a system can be described by the Lieb-Liniger Hamiltonian~\cite{Lieb1963May} 
\begin{align} \label{eq:HLL}
\hH = - \frac{\hbar^2}{2m}\sum_{j=1}^N \frac{\partial^2}{\partial x_j^2} + g \sum_{j<j'}^N \delta(x_j-x_{j'})\;,
\end{align}
where $m$ is particle mass and $g$ the coupling constant. The interaction regime is set by a dimensionless parameter $\gamma=mg/(\hbar^2\rho)$, known as the Lieb-Liniger parameter. 
The eigenstates $\Psi(\gamma,\bm{I})$ of the Hamiltonian~\eqref{eq:HLL} can be found exactly using Bethe ansatz and are uniquely labeled by the set of Bethe numbers $\bm{I}$ 
\cite{Lieb1963May,McGuire1964May}. Their corresponding eigenenergies, $E(\gamma,\bm{I}) = \frac{\hbar^2}{2m} \sum_j \lambda_j^2$, 
are expressed by the Lieb's quasi-momenta $\lambda_j$ that obey the Bethe equations
$\lambda_j L = 2\pi I_j-2\sum_{k=1}^{N} \arctan\big((\lambda_j-\lambda_{k})/(\gamma \rho)\big)$.
The exact solutions of the LL model can be used to examine the states emerging during the pumping procedure. We illustrate that in Fig.~\ref{fig:man_quasimomenta} showing Lieb's parameter $\gamma$ in the top panel, the corresponding energy of the system in the middle and the quasi-momenta of selected states at the bottom. As sketched in Fig.\ref{fig:man_quasimomenta}(a), the pumping protocol involves tuning the Lieb-Liniger parameter $\gamma$ via a control parameter. In the experiment of Ref.\cite{Kao2021Jan}, this was achieved by sweeping the magnetic field across a Feshbach resonance, allowing precise control of the three-dimensional scattering length and, via the confinement-induced resonance (CIR), fine-tuning of the one-dimensional interaction parameter $\gamma$. As the interaction strength is gradually increased, the ground state of an ideal Bose gas ($\gamma=0$) is adiabatically transformed into the ground state of TG gas ($\gamma\to+\infty$) with the energy $E_{\rm TG}/N=\frac{\pi^2\hbar^2(N^2-1)}{6m L^2}$. Subsequently, quench from strongly repulsive to strongly attractive interactions, $\gamma\! =\!+\!\infty \!\to\! \gamma\! =\!-\infty$, makes the gas pass from the ground to a highly-excited state. The resulting sTG gas is energetically far from the ground state of the attractive Lieb-Liniger system, which corresponds to McGuire's bright soliton with large and negative energy, $E_{MG}/N = -\frac{mg^2}{24\hbar^2}(N^2-1)$\cite{McGuire1964May}. A crucial characteristic of the sTG state is its metastability, as sTG state is very close to certain highly excited eigenstates, energetically separated from self-bound states.

Further, the adiabatic process can be extended by gradually tuning the interaction strength from strong attraction, $\gamma \to -\infty$, back to an excited state of the non-interacting Bose gas with the energy $E=4E_{\rm TG}$. This ends the first cycle, passing through energies marked with a turquoise line with circles in Fig.~\ref{fig:man_quasimomenta}(b). Additionally, adiabatic transitions to higher excited states are also feasible by continuing the process. This Letter focuses on the properties of states created during pumping at $\gamma=0$ and $|\gamma|=\infty$, labeled in Fig.~\ref{fig:man_quasimomenta} with $\ell$, with energies equal to $\ell^2 E_{\rm TG}$. During a single pumping cycle, the parameter $\ell$ is changed by $2$, so the number of cycles $n$ equals $2\ell$.

Throughout the pumping process, all changes in the gas states are smooth and continuous, even at the points of the resonance, where the interaction strength jumps from $\gamma=\infty \to \gamma=-\infty$. Nevertheless, the pumping connects eigenstates labeled by different sets of Bethe numbers and, in particular, states appearing at the same value of $\gamma$ but having different topological index $\ell$. These changes occur at the non-interacting points ($\gamma=0$) and in Fig.~\ref{fig:man_quasimomenta}(b) are marked by the change of the color of the line. The Bethe numbers change after crossing these points, but according to a clear pattern. In the first cycle (turquoise line in Fig.~\ref{fig:man_quasimomenta}(b)) the Bethe numbers are subsequent integers,  $\bm{I}^{(1)}=0,\pm1,\pm2,...$ in the second cycle they equal  $\bm{I}^{(2)}=0,\pm3,\pm6,...$,  in the third cycle $\bm{I}^{(3)}=0,\pm5,\pm10,...$ etc. This pattern is expected to occur for various system sizes \footnote{The pattern presented here is correct for odd numbers of atoms. For even ones, integers must be replaced by a half-integers, where $\bm{I}^{(1)}=\pm1/2,\pm3/2,...$, $\bm{I}^{(2)}=\pm3/2,\pm9/2,...$, and $\bm{I}^{(3)}=\pm5/2,\pm15/2,...$}. As a result, the Bethe numbers preserve the same structure throughout pumping, with extra rescaling occurring whenever interaction strengths pass through $\gamma=0$.

The pattern in the Bethe numbers translates to the characteristic distribution of Lieb's quasi-momenta, shown in Fig.~\ref{fig:man_quasimomenta}(c). In the ground state of the non-interacting gas, $\gamma=0$ and $\ell=0$, all quasi-momenta are zero ($\lambda_i = 0$), resembling a macroscopic occupation of the zero-momentum state in a Bose-Einstein condensate. For a finite interaction strength, $\gamma>0$, quasi-momenta separate from each other. On top of the resonance, $\gamma=+\infty$ and $\ell=0$, the Tonks-Girardeau gas is realized and Lieb's quasi-momenta follow a uniform distribution, exactly like the momenta in the ground state of the non-interacting fermions. 
%???
It remains unchanged during the quench but then, when interaction strength is gradually changed from $\gamma=-\infty$ to $\gamma=0$ the distribution further broadens. Eventually, for non-interacting bosons, the quasi-momenta distribution resembles again the distribution of momenta in the non-interacting fermions but $\ell=2$ larger spacing in momenta, namely the spacing equal to $2\cdot \left(2\pi/L\right)$. As the system passes through successive cycles, the quasi-momentum distribution undergoes continuous stretching. Thus, during the pumping process, the states at the special points $\gamma=0$ and $\gamma=\pm \infty$ feature quasi-momentum distributions equal to momenta of the ground state Fermi gas but scaled by a factor of $\ell$.
\begin{figure}
\centering
\includegraphics[trim= 1.5cm 0cm 0cm 0cm, width=\columnwidth]{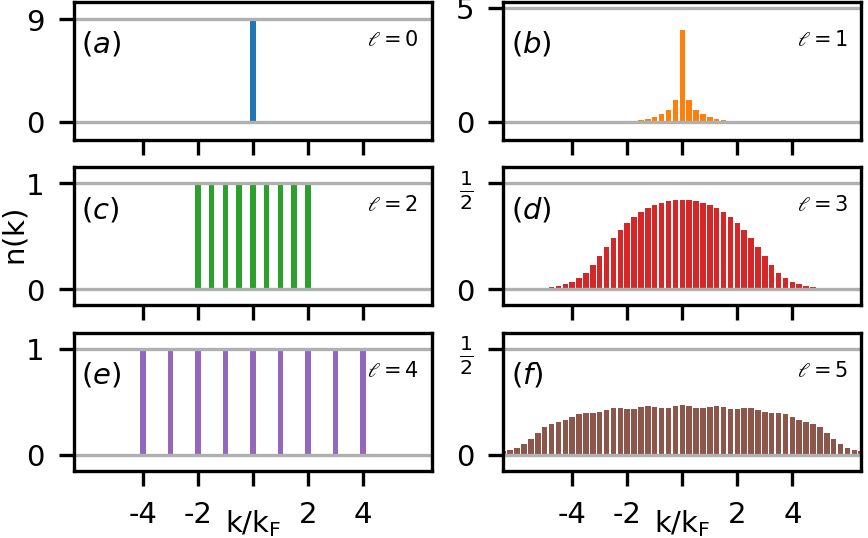}
\caption{
Momentum distributions of states appearing in the process of topological pumping at $\gamma=0$ (left column) and $\gamma=\pm\infty$ (right column) at the subsequent branches of the pumping labeled with $\ell=0,\,1,\,2,\,3,\,4,\,5$. Here, we considered $N=9$ atoms and we define Fermi momentum as $k_F=(N-1)\pi/L$.}
\label{fig:man_LL_momentum}
\end{figure}

The rescaling of the quasi-momentum distribution can also be observed in the wave function of the excited states.
Introducing operator $\mathcal{S}$, defined as $\mathcal{S}\Psi(x_1,...,x_N) = \prod_{i<j}\sign(x_j-x_i)\Psi(x_1,...,x_N)$, the state at successive stages of evolution for points $\ell=0,1,...$ can be represented as follows
\begin{align} 
\begin{split} \label{eq:LL_fock_states} 
\resizebox{0.9\columnwidth}{!}{
\xymatrix @C=1.5pc @R=.5pc{
&\underbrace{\ket{...,0,0,N,0,0,...}^B}_{\rm Ground\;state, \;\ell=0} \ar[r]^{g\to \infty} 
& \mathcal{S}\underbrace{\ket{...,1,1,1,...}^F}_{\rm Tonks-Girardeau, \;\ell=1}  \ar`r[d] `d[dl]<6pt> `l[ddll]<2pt>_{0 \leftarrow -\infty \leftarrow g} `d[dd]<-0pt> [ddl] & \\
& & &\\
&|...,\underbrace{1,0}_{\ell=2},1,0,1...\rangle^B 
\ar[r]<2pt>^{g\to \infty} & \mathcal{S}|...,\underbrace{1,0,0}_{\ell=3},1,0,0,1...\rangle^F \to \ldots
}
}
\end{split} \end{align}
where, $\ket{...}^{B(F)}$ is the bosonic (fermionic) Fock state, where single particle orbitals are the plane waves.

Finally, having discussed quasi-momenta, we show distributions of momenta. In the ideal gas, quasi-momenta and momenta coincide. Therefore, when $\gamma=0$ and $\ell=0,2,4,\cdots$, the momentum distribution looks like a comb, with peaks separated by $2\pi \ell/L$ -- see example in Fig.~\ref{fig:man_LL_momentum}(c,e). The momentum distribution is the same as for the ground state of non-interacting fermions, but in the box of size $L/\ell$. For completeness, we show in the right column of Fig.~\ref{fig:man_LL_momentum} the momentum distribution for states corresponding to $\ell=1, 3, 5$  in the TG regime ($\gamma=\infty$). These TG states have quasi-momenta $\bm{\lambda}$ equal to momenta of the Fermi gas (see Fig.~\ref{fig:man_quasimomenta}c), but the distributions of physical momenta are very different. In particular, in the ground state $\ell=1$ they exhibit a dominant mode at $k=0$.

\begin{figure}
\centering
\includegraphics[trim= 1.5cm 0cm 0cm 0cm, width=\columnwidth]{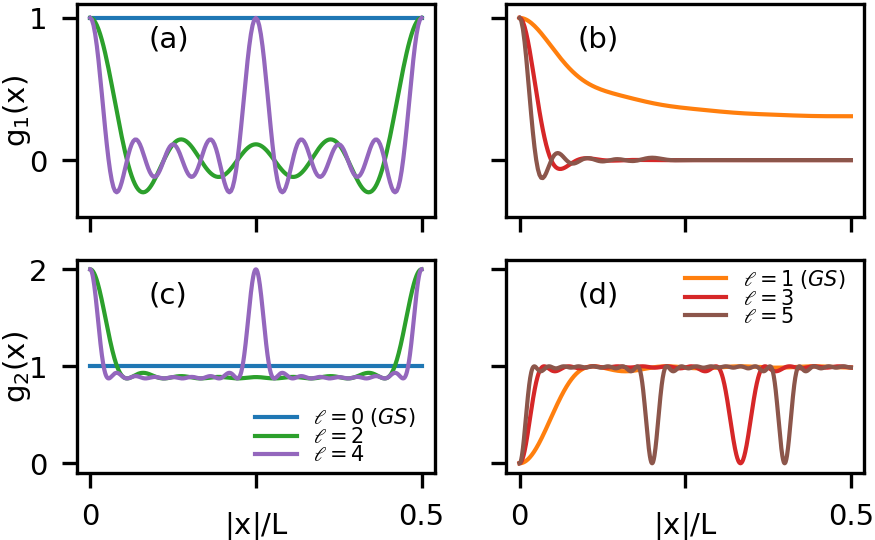}
\caption{
The first and the second order correlation functions of the states appearing in the process of topological pumping at $\gamma=0$ (left column, $\ell=0,\,2,\,4$) and $\gamma=\pm\infty$ (right column, $\ell=1,\,3,\,5$). Top panels: the first-order correlation $g_1$, the bottom: $g_2$. Results shown for $N=9$ atoms.}
\label{fig:man_LL_correlations}
\end{figure}

To investigate further the states emerging at $\gamma=0$ and $\gamma=\infty$ we show in Fig.~\ref{fig:man_LL_correlations} their first and second-order correlation functions, $g_1(x):=\meanv{\hat{\Psi}^{\dagger}(x)\hat{\Psi}(0)}/\rho$ (one-body density matrix, OBDM) and $g_2(x):=\meanv{\hat{\Psi}^{\dagger}(x)\hat{\Psi}^{\dagger}(0)\hat{\Psi}(0)\hat{\Psi}(x)}/\rho^2$ (pair distribution function), where $\hat{\Psi}(x)$ is bosonic field operator. Intuitively, the rescaling of the momentum distribution by a factor $\ell$ should be accompanied by an inverse rescaling of certain spatial properties of the system. The comb-like structure of the quasi-momentum distribution of the ideal gas results in a self-similarity property of the correlation functions (left column in Fig.~\ref{fig:man_LL_correlations}). For example, the correlation functions for $\ell=4$ (violet line) can be obtained from the ones for $\ell=2$ (green line) by squeezing by a factor of $2$ and then doubling the number of resulting curves.
 
We focus on the connection between the excited states in ideal bosons, $\gamma=0$, and the ground state in ideal fermionic gases by comparing their coherence properties (quantified with $g_1$). Both systems have the same momentum distribution, and consequently, they share the same one-body density matrix $g_1(x)$, which is related to $n(k)$ via the Fourier transform,
\begin{equation}
g_{1,{\rm even}\;\ell}(x) = \frac{1}{N}\sum_{j=-\floor{N/2}}^{\floor{N/2}} 
%e^{i 2\pi j \ell x /L}  
e^{\frac{i 2\pi j \ell x}{L}}  
    =\frac{\sin(N\pi \ell x/L)}{N \sin(\pi \ell x/L)}.
\label{Eq:G1:even}
\end{equation}
In the thermodynamic limit, $g_{1,{\rm even}\;\ell}(x) = \sin(\ell k_F x)/(\ell k_F x)$, revealing that the coherence decays as $\propto 1/ \ell x$. Moreover, OBDM exhibits the Friedel oscillations appearing at multiples of $1/(\ell\rho)$, a behavior which is highly unusual for bosons, which typically have a positive and non-oscillatory OBDM in the ground state. It is instructive to analyze the short-length expansion of the OBDM in $\gamma=0$ and $\gamma=\infty$ states
\begin{align*}
\begin{cases}
g_{1,even\;\ell}(x) = 1-\frac{(\pi \ell)^2(1-\frac{1}{N^2})}{6} (\rho x)^2 + O(x^4)
    %+    \frac{(\pi \ell)^4}{120}(\rho x)^4
    \\
g_{1,odd\;\ell}(x)\!=\!1\!-\!\frac{\!\!(\pi\ell)^2(1-\frac{1}{N^2})(\rho x)^2\!\!}{6}\!+\!\frac{\!\!(\pi \ell)^2(1-\frac{1}{N^2})|\rho x|^3\!\!}{9}  
+O(x^4)
%+    \frac{(\pi \ell)^4}{120}(\rho x)^4
\end{cases}
\end{align*}
which comes in terms of analytic expansion in powers of $x$ and non-analytic terms containing $|x|$~\cite{PhysRevE.74.021105}. The $n$-th coefficient of the analytic expansion is proportional to the $n$-th moment of the momentum distribution, $c_n = \frac{i^n}{\rho^n} \int k^n\frac{n(k)}{\rho} \frac{dk}{2\pi}$. In particular, the quadratic term is proportional to the kinetic energy which for $\gamma=0$ and $\gamma=\infty$ coincides with the total energy. Instead, the coefficient in the non-analytic $|x|^3$ term is proportional to the derivative of the equation of state with respect to $\gamma$ as first established by Olshanii and Dunko\cite{PhysRevLett.91.090401} and later identified as Tan's contact. As the energies of the considered states are given by the Fermi energy scaled by $\ell^2$, both coefficients are equal to these in an ideal Fermi gas, scaled by $\ell^2$.

We complement the discussion of correlations by showing the second-order correlation functions in Fig.~\ref{fig:man_LL_correlations}c) and d).  
The $g_{2}$ correlation functions in the TG gas (Fig.~\ref{fig:man_LL_correlations} d) are the same as in a system of fermions with densities $\ell N/L$, in particular dropping to zero at points separated by $L/\ell$. On the other hand, the spatial correlations in the ideal Bose gas, Fig.~\ref{fig:man_LL_correlations} c), exhibit bunching. Due to above-mentioned self-similarity, a highly excited state ($\ell \gg 1$) will show a periodic structure with increased probability of finding an atom at positions that are multipoles of $L/\ell$. 

%%%%%%%%%%%%%%%%%%%%%
\ssec{Conclusions}%%%
%%%%%%%%%%%%%%%%%%%%%
Our theoretical study is inspired by the experimental results on topological pumping of a trapped quasi-1D bosonic gas \cite{Kao2021Jan}. We focus on a strictly 1D homogeneous case to make use of the analytical results for the Lieb-Liniger model. We show that in this case the protocol of the topological pumping would lead to a highly correlated state, even for an ideal gas. 

The {\it momentum distribution of the non-interacting gas ($\gamma=0$)}, after the $n$ pump cycles, is exactly the same as that in the ground state of fermions in a box of size $L/(2n)$. If the pumping continues to Tonks-Girardeau regime, then its {\it spatial distribution and all density-related correlations at $\gamma=\infty$} become exactly {\it the same as for fermions in a ground state, but in a shorter box}. 

As a direct consequence, the one-body density matrix of the state of the non-interacting gas after $n$-cycles has sign-changing Friedel oscillations, resembling the behavior in ideal fermions, despite the pair distribution function showing a bosonic enhancement, i.e. with $g_2(x=0)=2$. Also, we analyze the short-range expansion of the one-body density matrix of the excited states emerging in pumping at $\gamma=0$ and $\gamma=\infty$, and argue that the coefficient of $x^2$ is proportional to the kinetic energy, while $|x|^3$ term corresponds to the Tan's contact -- results known for the ground-states, but apparently emerging also in the gas excited due to the pumping. Our analysis reveals the interesting states that can be generated in the currently used experimental technique, and opens a pathway to study their properties and potential applications.

\acknowledgments

M. M., and K.P. acknowledge support from the (Polish) National Science Center Grant No. 2019/34/E/ST2/00289. Center for Theoretical Physics of the Polish Academy of Sciences is a member of the National Laboratory of Atomic, Molecular, and Optical Physics (KL FAMO). 
G.E.A. acknowledges the support of the Spanish Ministry of Science and Innovation (MCIN/AEI/10.13039/501100011033, grant PID2023-147469NB-C21), the Generalitat de Catalunya (grant 2021 SGR 01411) and {Barcelona Supercomputing Center MareNostrum} ({FI-2025-1-0020}).
B. J-D acknowledges support from Grant PID2023-147475NB-I00 funded by MICIU/AEI/10.13039/501100011033 and FEDER, UE, and grant 2021 SGR 01095 from Generalitat de Catalunya, and by Project CEX2019-000918-M of ICCUB (Unidad de Excelencia María de Maeztu).

\bibliographystyle{apsrev4-1}
\bibliography{bibliography.bib}

\onecolumngrid
\section{Supplemental material}
In the limit of noninteracting bosons, the single particle density matrix $G_1(x) = \langle \hPsid(x) \hPsi(0) \rangle$ is simply a Fourier transform of the momentum distribution
\begin{align}
    G_{1,even\;\ell}(x)  &= \frac{1}{L}\sum_{j=-N/2}^{N/2} e^{i \frac{2\pi \ell}{L} j  x /L} = \frac{e^{i\frac{\pi(N-1)}{L}}}{L} \frac{1-\big(e^{i\frac{2 \pi \ell}{L} x}\big)^N}{1-e^{i\frac{2 \pi \ell}{L} x}} = \frac{\sin(N\pi \ell x/L)}{L \sin(\pi \ell x/L)}
\end{align}
and in the limit of short distances between atoms $x \to 0$ and for large numbers of atoms $N \to \infty$ can be approximated by
\begin{align}
    G_{1,even\;\ell}(x)/\rho &\xrightarrow[N \gg 1]{x \ll L} \sum_{j=-N/2}^{N/2} \big( 1 - \frac{(2\pi \ell j)^2}{2L^2}x^2 + \frac{(2\pi \ell j)^4}{4! L^4}x^4\big) 
    =1-\frac{(\pi \ell)^2}{6} (\rho x)^2 
    +
    \frac{(\pi \ell)^4}{120}(\rho x)^4
\end{align}
In the limit of strong contact repulsion $\gamma \to \infty$ single particle correlation function takes the form of
\begin{align}
    G_{1,odd\;\ell}(x)
     = \frac{2}{N L^N}\sum_{j=-N_h}^{0}\Re \bigg( \exp(i \frac{2\pi \ell j}{L}x) H_{\ell}(x,j) \bigg)
\end{align}
where:
\begin{align}
    H_{\ell}(x,j) = \sum_{P}\rm{sign}(P)\prod_{k=-N_h,k\neq j}^{N_h} F_{\ell}(x,j,Pk)
\end{align}
with
\begin{align}
    F_{\ell}(x,j,j') = \begin{cases}
        L-2x,\;\; if\; j=j'\\
        -\frac{2e^{i\frac{2\pi \ell}{L}(j-j')x}-2 }{i\frac{2\pi \ell}{L}(j-j')},\;\; if\; j \neq j'.
    \end{cases}
\end{align}
In the limit of $x \to 0$ and $N \to \infty$ with constant density
\begin{align}
    F_{\ell}(x,j,j') \xrightarrow{x\to 0} \begin{cases}
        L-2x,\;\; if\; j=j'\\
        -x-i\frac{\pi \ell}{L}(j-j')x^2+\frac{\pi \ell}{3L}(j-j')x^3 +O(x^4),\;\; if\; j \neq j'.
    \end{cases}
\end{align}
It can be seen that in this regime in the excited states where $\ell > 1$, all coefficients $\pi$ are replaced by $\pi \ell$. After all, one can expect that the single particle density matrix on short distances takes a form
\begin{align}
    G_{1,odd\;\ell}(x)/\rho \xrightarrow[N \gg 1]{x \ll L} 1 - \frac{(\pi \ell)^2}{6} (\rho x)^2 + \frac{(\pi \ell)^2}{9} |\rho x|^3 
\end{align}
\end{document}